\documentclass[10pt]{article}
\usepackage{geometry}                
\usepackage[parfill]{parskip}
\usepackage{amssymb}

\usepackage{amsmath}

\title{\bf Generalised virial theorems in Classical and Quantum Mechanics}

\author{C. V. Sukumar \\{\em Wadham College,}\\{\em University of Oxford, Oxford OX1 3PN, U.K. }}

\begin{document}
\maketitle

\begin{abstract}

Generalizations of the virial theorem in Classical Mechanics and Quantum Mechanics are examined. It is shown that the generalized virial theorem in Quantum Mechanics leads to certain relations between matrix elements. The differences between the generalizations in Classical and Quantum Mechanics are identified. Some results arising from the radial Schr\"{o}dinger equation in Quantum Mechanics are discussed. It is also shown that the generalizations of the virial theorem may be extended to arbitrary number of dimensions.

\end{abstract}

\section{Introduction}

The virial theorem in Classical Mechanics provides a relation between the time averages of the kinetic energy and of ${\bf r.F}$ for periodic orbits of a system of particles subject to forces ${\bf F=-\nabla}V$ where $V$ is the potential energy of the system. The classical virial theorem has proved useful not only in Astrophysics and Cosmology which are concerned with bounded motion in a gravitational potential but also in Thermodynamics because the average kinetic energy may be related to the temperature. The virial theorem in Classical Mechanics also has an analogue in non-relativistic Quantum Mechanics and provides a relation between the expectation values of the operators representing the kinetic energy and ${\bf r.\nabla}V$ when the expectation values are evaluated in the eigenstates of the potential $V$ and the square of the angular momentum $l^2$ in Classical Mechanics is replaced by the eigenvalues of the $L^2$ operator in Quantum Mechanics. Quigg and Rosner (1979) have considered some generalizations of the virial theorem in Quantum Mechanics and explored the consequences for the non-relativistic analysis of quarkonium spectra. The so-called  Kramer's relation (Messiah 1966) between the matrix elements of various operators arising in the study of spectroscopy is a  particular example of the generalization of the virial theorem. In this report we show that the analysis of Quigg and Rosner may be placed in an even more general formulation. We use the feature that the square of the bound state radial wavefunction in a spherically symmetric potential satisfies a third order differential equation and the bounded solution to this equation may be used as a probability distribution to define the expectation values in Quantum Mechanics. We show that the generalized virial theorems in Classical and Quantum Mechanics differ in some respects. We examine the similarities and differences between the generalizations in Classical and Quantum Mechanics. We also discuss how the generalizations may be extended to an arbitrary number of spatial dimensions. 

\section{Generalized virial theorem in Quantum Mechanics}

The Schr\"{o}dinger equation for the bound states of a potential $V$ in $3$ dimensions may be given in the form
\begin{equation}
-\frac{\hbar^2}{2M}\ \Big[\frac{\partial^2}{\partial r^2} + \frac{2}{r} \frac{\partial}{\partial r} + \frac{{\bf L}^2}{r^2} \Big] \Psi = \big[E\ -\ V \big]\ \Psi\ . \label{}
\end{equation}
For a spherically symmetric potential the radial and angular parts of the normalized wavefunction $\Psi$ may be separated in the form $\Psi = R/r \operatorname{Y}_{lm}(\theta,\phi)$ and the eigenstates may be expressed in terms of radial eigenfunctions which are solutions to the radial equation 
\begin{equation}
\frac{\partial^2 R_{nl}}{\partial\rho^2} = Q\ R_{nl} ,\qquad Q = -2 \Big(\epsilon - v - \frac{l(l+1)}{2\rho^2}\Big), \label{eq:N1}
\end{equation}
where $\rho$ is a dimensionless radial coordinate defined by $r=\rho a$ where $a$ is a scaling length associated with the potential energy $V(r)$ of the Hamiltonian $H$. $\epsilon$ is the scaled energy measured in units of energy defined in terms of the scaling length and the mass $M$ and $v(\rho)$ is the scaled potential. The full set of defining relations is:
\begin{align}
|n,l \rangle &= R_{nl}(\rho) \operatorname{Y}_{lm}(\theta,\phi) ,\qquad {\bf L}^2 \operatorname{Y}_{lm}(\theta,\phi) = l(l+1)\operatorname{Y}_{lm}(\theta,\phi),\qquad 
(h_0  - \epsilon) |n, l \rangle = 0  \label{eq:N4}\\
 h_0 - \epsilon &= \frac{Ma^2}{\hbar^2}  (H - E) = - \frac{1}{2} \frac{\partial^2}{\partial \rho^2} + \frac{{\bf L}^2}{2\rho^2} +  v - \epsilon ,\qquad  v = \frac{Ma^2}{\hbar^2} V ,\qquad \epsilon = \frac{Ma^2}{\hbar^2} E .\label{eq:N2}
\end{align}

Let $P_{nl}=R_{nl}^2$ be the probability density associated with the radial Schr\"{o}dinger equation. Using the notation that a dot denotes a derivative with respect to $\rho$ and suppressing the suffixes $n$ and $l$ it can be established that
\begin{align}
{\dot P} &= 2 R {\dot R} \notag \\
{\ddot P} &= 2(R{\ddot R} + {\dot R}^2) = 2(QR^2 + {\dot R}^2) = 2(QP + {\dot R}^2) \notag \\
{\dddot P} &= 2({\dot Q} P + Q{\dot P} + 2{\dot R} {\ddot R}) = 2({\dot Q} P + Q{\dot P} + 4 R {\dot R} Q) = 2({\dot Q} P + 2Q{\dot P} ) , \label{eq:N5}
\end{align}
which is a third order differential equation satisfied by $P$. This is to be expected since a second order differential equation has two linearly independent solutions and there are three possible products of the two independent solutions one of which is $P$. From (\ref{eq:N5}) it follows that for any function $f(\rho)$ we have
\begin{equation}
\int_0^{\infty}  f \Big(\frac{{\dddot P}}{2} - {\dot Q} P - 2Q{\dot P}\Big) {\text{d}}\rho = 0 .\label{}
\end{equation}
For any function $f(\rho)$ which grows less fast than $P^{-1}$ as $\rho\rightarrow \infty$ integration by parts and use of the limiting relations given by
\begin{align}
\operatorname{Lt}_{\rho\rightarrow 0} &P = C_{nl}^2 \rho^{2l+2},\qquad \operatorname{Lt}_{\rho\rightarrow 0} f = b \rho^q, \qquad  \operatorname{Lt}_{\rho\rightarrow 0} Q = \frac{l(l+1)}{\rho^2},\qquad \operatorname{Lt}_{\rho\rightarrow\infty} (Pf) = 0 \\
\operatorname{Lt}_{\rho\rightarrow 0} &\Big[\frac{1}{2} \Big( f{\ddot P} - {\dot f} {\dot P}  + {\ddot f} P\Big) - 2 f PQ\Big] =  C_{nl}^2 b \Big[(1-q) \Big(l+1 - \frac{q}{2}\Big)\Big] \rho^{q+2l} ,\label{}
\end{align}
leads to the establishment of the relation
\begin{equation}
\int_0^{\infty} P \Big({\dot Q} f + 2Q {\dot f} - \frac{1}{2} {\dddot f} \Big) {\text{d}}\rho = C_{nl}^2  b (2l+1)^2 \delta_{q,-2l},  q\ge (-2l) .\label{eq:N6}
\end{equation}
Since integrals weighted by the probability distribution $P$ are related to the concept of expectation values in Quantum Mechanics the relation we have derived may be written in the form
\begin{equation}
\left\langle \frac{1}{f} \frac{\partial(f^2 Q)}{\partial\rho}\right\rangle - \frac{1}{2} \left\langle \frac{\partial^3f}{\partial\rho^3}\right\rangle = C_{nl}^2  b (2l+1)^2 \delta_{q,-2l} ,\label{eq:N7}
\end{equation}
which is a generalization of Kramer's relation (Messiah, Kramer, Quigg and Rossner). The particular choice $f=\rho^j$ leads to the relation
\begin{equation}
4j \Big[-\epsilon\left\langle \rho^{j-1}\right\rangle + \left\langle v \rho^{j-1}\right\rangle \Big] + 2 \left\langle \rho^j \frac{\partial v}{\partial\rho}\right\rangle + \frac{(j-1)}{2} (2l+j) (2l+2-j) \left\langle \rho^{j-3}\right\rangle = C_{nl}^2  (2l+1)^2 \delta_{j,-2l} ,\label{eq:N8}
\end{equation}
which is valid for all real values of $j\ge (-2l)$. This result further simplifies if at least one of the terms in (\ref{eq:N8}) vanishes. This is the case if $j=0,1,2l+2,-2l$. We examine these special cases first.
 
${\bullet}$ The choice $j=0$ leads to the result
\begin{equation}
V_{\text{eff}} \equiv v + \frac{l(l+1)}{2\rho^2} , \left\langle nl\middle| \frac{\partial V_{\text{eff}}}{\partial\rho}\middle|nl\right\rangle = \frac{C_{nl}^2}{2} \delta_{0,l}\qquad , \label{eq:N9}
\end{equation}
which shows that the expectation value of the effective force is zero for the states which are not spherically symmetric. For $l=0$ the expectation value of the force is only proportional to the square of the eigenfunction at $\rho=0$. 

The probability density $|\Psi_{n,0}(0)|^2$ is of interest, for example, in the leptonic decays of massive neutral vector mesons $V^0$ which are $^3{\text{S}}_1$ bound states of a quark and an anti-quark. The decay width (Quigg and Rossner p186) is given by
\begin{equation}
\Gamma\big(V^0\rightarrow l^+l^-\big) = 16\pi \frac{\hbar^3}{c} \Big(\frac{\alpha_{\text{e}} {\text e}_{\text{q}}}{M_{\text{V}}}\Big)^2 |\Psi(0)|^2 = 4 \frac{c\hbar}{a} \Big(\frac{\hbar \alpha_{\text{e}} {\text{e}}_{\text{q}}}{M_{\text{V}} c a}\Big)^2 C_{n0}^2  ,\label{eq:N10}
\end{equation}
where $e_{\text{q}}$ is the charge of the quark in units of the electron charge, $\alpha_{\text{e}}$ is the fine structure constant, $c$ is the speed of light and $M_{\text{V}}$ is the mass of the vector meson. It is evident from eqs.  (\ref{eq:N9}) and (\ref{eq:N10}) that the decay width is directly proportional to the expectation value of the force in a spherically symmetric eigenstate of the vector meson.

${\bullet}$ The choice $j=1$ leads to the virial theorem
\begin{equation}
\left\langle \rho \frac{\partial v}{\partial \rho}\right\rangle = 2 \left\langle \epsilon-v\right\rangle = 2  \left\langle T \right\rangle  ,\label{eq:N12}
\end{equation}
where $T$ is the kinetic energy. The virial theorem has an analogue in Classical Mechanics when the quantum expectation values are replaced by time averages over the period of a bound orbit in Classical Mechanics. 

Higher values of $j$ lead to generalizations of the virial theorem. It will be shown later that the classical and quantum results agree for $j=1$ and $j=2$ and that for other values of $j$ there is a discrepancy. The analysis in  Classical Mechanics does not yield a term corresponding to $\big<{\dddot f}\big>$ in the Quantum analysis. This additional term in the Quantum analysis gives zero for the cases $j=1,2$ but gives non-vanishing values for other values of $j$. The generalization of the virial theorem in Classical Mechanics will be examined in a later section.

${\bullet}$ The choice $j=2l+2$ leads to 
\begin{equation}
\left\langle nl\middle|\rho^{2l+2} \frac{\partial v}{\partial \rho}\middle|nl \right\rangle = 4(l+1) \left\langle nl\middle|\rho^{2l+1} (\epsilon_{nl} - v) \middle|nl\right\rangle .\label{eq:N13}
\end{equation}

${\bullet}$ The choice $j=-2l$ leads to
\begin{equation}
8l \left\langle nl\middle|\rho^{-2l-1}(\epsilon-v)\middle|nl \right\rangle + 2\left\langle nl\middle| \rho^{-2l} \frac{\partial v}{\partial\rho}\middle|nl\right\rangle = C_{nl}^2  (2l+1)^2 .\label{eq:N14}
\end{equation}

In general if $j$ is not equal to one of the special values considered above then (\ref{eq:N8}) provides a connection between various matrix elements. For positive definite values of $j$ the term on the right hand side of (\ref{eq:N8}) vanishes and the relation may be given in the form
\begin{equation}
\left\langle \rho^j \frac{\partial v}{\partial \rho}\right\rangle = 2j \left\langle (\epsilon_{nl} - v) \rho^{j-1}\right\rangle - \frac{(j-1)}{4} (2l+j) (2l+2-j) \left\langle\rho^{j-3}\right\rangle\quad\text{for $j>0$} .\label{eq:n11}
\end{equation}
This general result can be elucidated by considering some illustrative examples.

${\bullet}$ For example, the choice $j=2$ leads to 
\begin{equation}
\left\langle \rho^2 \frac{\partial v}{\partial \rho}\right\rangle  = 4 \left\langle\rho (\epsilon - v)\right\rangle  - \left\langle \frac{l(l+1)}{\rho}\right\rangle .\label{}
\end{equation}

${\bullet}$ The choice $j=3$ leads to 
\begin{equation}
\left\langle \rho^3 \frac{\partial v}{\partial \rho}\right\rangle  = 6 \left\langle\rho^2 (\epsilon - v) \right\rangle  - \frac{1}{2} (2l-1)(2l+3) .\label{}
\end{equation}

\subsection{Moments of the kinetic energy operator}

We next examine some general results arising from the commutation relations in Quantum Mechanics with a view to studying the expectation values of powers of the kinetic-energy operator. Using the commutation relation $[ {\bf p}, V] = -{\text{i}}\hbar {\bf \nabla} V$ it can be established that
\begin{equation}
[ H,V] = -\frac{\hbar^2}{2M} \Big(\nabla^2 V + 2 {\bf \nabla} V. {\bf \nabla} \Big) .\label{eq:g0}
\end{equation}
Using this result it may be shown that for the scaled Hamiltonian $h_0$ and the scaled potential ~$v$
\begin{align}
h_0 v R &= \epsilon v R  - \frac{1}{2} {\ddot v} R - {\dot v} {\dot R} \label{eq:g2} \\
\left\langle v^j [h_0,v] \right\rangle &= -\frac{1}{2} \left\langle \Big(v^j {\ddot v} - \frac{\partial}{\partial\rho}\big( v^j {\dot v}\big)\Big)\right\rangle = \frac{j}{2} \left\langle v^{j-1} {\dot v}.{\dot v} \right\rangle \\
\left\langle v h_0 v \right\rangle &=  \epsilon\left\langle v^2\right\rangle + \frac{1}{2} \left\langle{\dot v}^2 \right\rangle .\label{eq:g3}
\end{align}
Using (\ref{eq:g2}) and its adjoint it may be shown that
\begin{align}
\left\langle R\middle| v h_0^2 v\middle|R\right\rangle &= \left\langle R\middle|\big[ \epsilon^2 v^2 - \epsilon v {\ddot v} + \frac{1}{4}{\ddot v}^2 \big]\middle|R\right\rangle + \left\langle {\dot R}\middle| {\dot v}^2\middle|{\dot R}\right\rangle - 2 \left\langle R\middle| {\dot v} \epsilon v \middle|{\dot  R}\right\rangle + \left\langle R\middle| {\dot v} {\ddot v}\middle|{\dot R}\right\rangle {\notag}\\
&= \left\langle R\middle|\big[ \epsilon^2 v^2 - \epsilon v {\ddot v} + \frac{1}{4}{\ddot v}^2 \big]\middle|R\right\rangle  - \left\langle R\middle|{\dot v}^2\middle|{\ddot R}\right\rangle + \epsilon \left\langle R\middle| \Big(\frac{\partial}{\partial\rho}({\dot v}v)\Big)\middle|R\right\rangle - \left\langle R\middle| {\dot v} {\ddot v}\middle|{\dot R}\right\rangle {\notag}\\
&= \left\langle R\middle|\Big[ \epsilon^2 v^2 + \epsilon {\dot v}^2 + \frac{1}{4} {\ddot v}^2  + 2{\dot v}^2 \Big(\epsilon-v - \frac{l(l+1)}{2\rho^2}\Big)\Big]\middle|R\right\rangle +\frac{1}{2}\left\langle R\middle| \Big(\frac{\partial}{\partial\rho} {\dot v}{\ddot v}\Big)\middle|R\right\rangle {\notag} \\
&= \left\langle R\middle| \Big[ \epsilon^2 v^2 + 3\epsilon {\dot v}^2 +\frac{3}{4} {\ddot v}^2 +\frac{1}{2} {\dot v} {\dddot v} -2 v {\dot v}^2 -l(l+1) {\dot v}^2 \frac{1}{\rho^2}\Big]\middle|R\right\rangle .\label{eq:g4}
\end{align}

Representing the kinetic energy by $T$ and using eqs. (\ref{eq:g3}) and (\ref{eq:g4}) it can be shown that
\begin{align}
\left\langle T \right\rangle &= \left\langle (h_0 - v) \right\rangle = \epsilon - \left\langle v \right\rangle \label{eq:g5}\\
\left\langle T^2 \right\rangle &= \left\langle (h_0 - v)^2 \right\rangle = \epsilon^2 - 2\epsilon \left\langle v \right\rangle  + \left\langle v^2 \right\rangle = \left\langle (\epsilon - v)^2 \right\rangle \label{eq:g6} \\
\left\langle T^3 \right\rangle &= \left\langle (h_0 - v)^3 \right\rangle = \epsilon^3 - 3\epsilon^2\left\langle v \right\rangle  + 3\epsilon \left\langle v^2 \right\rangle + \left\langle v [h_0,v]\right\rangle - \left\langle v^3 \right\rangle \notag \\
&= \left\langle (\epsilon-v)^3\right\rangle + \frac{1}{2} \left\langle \Big(\frac{\partial v}{\partial \rho}\Big)^2 \right\rangle\label{eq:g7}\\
\left\langle T^4 \right\rangle &= \left\langle (h_0 -v)^4 \right\rangle \notag \\
&= \epsilon^4 - 4 \epsilon^3 \left\langle v\right\rangle + 3 \epsilon^2 \left\langle v^2\right\rangle + 2\epsilon \left\langle v h_0 v\right\rangle +\left\langle vh_0^2 v\right\rangle - 2\epsilon \left\langle v^3\right\rangle -2 \left\langle v^2 h_0 v\right\rangle + \left\langle v^4\right\rangle \notag \\
&= \left\langle (\epsilon-v)^4 \right\rangle + 2\epsilon \left\langle v (h_0 -\epsilon) v\right\rangle + \left\langle v (h_0^2 - \epsilon^2) \right\rangle - 2 \left\langle v^2 (h_0-\epsilon) v\right\rangle \notag \\
&= \left\langle (\epsilon-v)^4 \right\rangle + 4 \left\langle \Big(\epsilon-v -\frac{l(l+1)}{4\rho^2}\Big) {\dot v}^2\right\rangle + \frac{3}{4} \left\langle {\ddot v}^2\right\rangle + \frac{1}{2} \left\langle {\dot v} {\dddot v}\right\rangle .  \label{eq:g8}
\end{align}

We can illustrate the results from this section using the example of power law potentials and some exactly solvable problems.

\section{Power law potentials}

We consider the power law potential $v=A\rho^m/2$. If $f=\rho^j$ the relation in (\ref{eq:N8}) may be given in the form
\begin{equation}
-4j\epsilon\left\langle\rho^{j-1}\right\rangle + (2j+m)A \left\langle \rho^{j+m-1}\right\rangle + \frac{(j-1)}{2} (2l+j)(2l+2-j) \left\langle\rho^{j-3}\right\rangle = C_{nl}^2 (2l+1)^2 \delta_{j,-2l} . \label{eq:P0}
\end{equation}
In general this equation relates three different matrix elements. However, for power law potentials  the relations simplify not only when $j=0,1,-2l$ or $2l+2$ but also if $j=-m/2$. For these special values of $j$ at least one of the terms in the equation vanishes and a relation connecting two different matrix elements emerges:
\begin{alignat}{2}
j=0, l=0 &\Longrightarrow & A \left\langle n0\middle|\rho^{m-1}\middle|n0\right\rangle &= \frac{C_{n0}^2}{m} \label{eq:P1}\\
j=0, l>0 &\Longrightarrow & A \left\langle nl\middle|\rho^{m-1}\middle|nl\right\rangle &= \frac{2}{m} \left\langle nl\middle|\frac{l(l+1)}{\rho^3}\middle|nl\right\rangle \label{eq:p1}\\
j=1 &\Longrightarrow & \left\langle nl\middle|v\middle|nl\right\rangle &= A \left\langle nl\middle|\frac{\rho^m}{2}\middle|nl\right\rangle = \frac{2}{2+m} \epsilon_{nl} \label{eq:P2}\\
j=-\frac{m}{2}, l>\frac{m}{4} &\Longrightarrow & \left\langle nl\middle|\rho^{-m/2-1}\middle|nl\right\rangle &= \frac{m+2}{32m\epsilon_{nl}} (4l-m)(4l+4+m) \left\langle nl\middle|\rho^{-m/2-3}\middle|nl\right\rangle \label{eq:P3}\\
j=-2l, l>0 &\Longrightarrow & 8l\epsilon_{nl} \left\langle nl\middle|\rho^{-2l-1}\middle|nl\right\rangle &= (4l-m)A \left\langle nl\middle|\rho^{m-2l-1}\middle|nl\right\rangle + C_{nl}^2 (2l+1)^2\label{eq:P4} \\
j=2l+2 &\Longrightarrow & A \left\langle nl|\rho^{2l+1+m}|nl\right\rangle &= \frac{8(l+1)\epsilon_{nl}}{4l+m+4} \left\langle nl|\rho^{2l+1}|nl\right\rangle .\label{eq:P5}
\end{alignat}
(\ref{eq:P2}) is a statement of the virial theorem for power law potentials. For spherically symmetric states (\ref{eq:P1}), (\ref{eq:P2}) and (\ref{eq:P5}) are applicable if $m>0$. For $l\geq 1$ states, depending on the value of $m$ and $l$ four or five of the relations given in (\ref{eq:p1})--(\ref{eq:P5}) are applicable. In general
\begin{equation}
-4j\epsilon\left\langle\rho^{j-1}\right\rangle + (2j+m)A \left\langle\rho^{j+m-1}\right\rangle + \frac{(j-1)}{2} (2l+j)(2l+2-j) \left\langle\rho^{j-3}\right\rangle  = 0  , j>0 . \label{eq:P6}
\end{equation}
This relation provides a recursive scheme for evaluating the matrix elements of $\rho^j$. We now consider some exactly solvable cases of power law potentials.

\subsection{Simple harmonic oscillator}

For the oscillator potential $v=\rho^2/2$ the relation given in (\ref{eq:P1}) leads to
\begin{equation}
\left\langle nl\middle| \rho \middle| nl\right\rangle =   \frac{|\Psi_{nl}(0)|^2}{2} \delta_{0,l} + \left\langle nl\middle| \frac{l(l+1)}{\rho^3} \middle| nl \right\rangle .\label{eq:M1}
\end{equation}
For $l=0$ the normalized radial wavefunctions are given by
\begin{align}
\rho \Psi_{n0} \equiv R_{n0} &= \frac{1}{\pi^{1/4}} \frac{1}{\sqrt{2^{2n} (2n+1)!}} \exp\Big(-\frac{\rho^2}{2}\Big) H_{2n+1}(\rho)\\
\operatorname{Lt}_{\rho\to 0} H_{2n+1}(\rho) &= (-)^n \frac{(2n+2)!}{(n+1)!} \rho  , \label{}
\end{align}
where $\operatorname{H}_m(\rho)$ are Hermite polynomials. In this case (\ref{eq:M1}) leads to 
\begin{align}
\left\langle n0 | \rho |n0\right\rangle &= \frac{1}{\sqrt{\pi}} \frac{1}{2^{2n} (2n+1)!} I_{nn} \label{eq:M2}\\
I_{nn} &=\int_0^{\infty} \exp(-\rho^2) \big[\operatorname{H}_{2n+1}(\rho)\big]^2 \rho {\text{d}}\rho = \frac{1}{2} \Big[\frac{(2n+2)!}{(n+1)!}\Big]^2 .\label{eq:M3}
\end{align}
This may be verified using Mehler's formula (Wikipedia) for a generating function for the Hermite polynomials given in the form
\begin{equation}
\sum_{n=0}^{\infty} \operatorname{H}_n(x) \operatorname{H}_n(y) \frac{1}{n!} \frac{u^n}{2^n} = \frac{1}{\sqrt{1-u^2}} \exp\Big[\frac{2u}{1+u} xy - \frac{u^2}{1-u^2} (x-y)^2\Big]  ,\label{}
\end{equation}
which can be used to show that
\begin{equation}
\sum_{n=0}^{\infty} \frac{\operatorname{H}_{2n+1}^2(\rho)}{(2n+1)!} \frac{u^{2n+1}}{2^{2n+1}} = \frac{1}{2} \frac{1}{\sqrt{1-u^2}} \Big(\exp\Big[\frac{2u}{1+u} \rho^2\Big] - \exp\Big[-\frac{2u}{1-u} \rho^2\Big]\Big)  .\label{}
\end{equation}
This leads to the integral expression
\begin{align}
\int_{0}^{\infty}\sum_{n=0}^{\infty} \frac{\operatorname{H}_{2n+1}^2(\rho)}{(2n+1)!} \frac{u^{2n+1}}{2^{2n+1}} \exp(-\rho^2) \rho^{2k-1} {\text{d}}\rho &= \frac{1}{4} \frac{1}{\sqrt{1-u^2}} \int_{0}^{\infty} \Big(\exp(-\beta z) - \exp\big(-\frac{z}{\beta}\big)\Big) z^{k-1} {\text{d}}z \notag \\
 \beta &= \frac{1-u}{1+u} ,\qquad z = \rho^2  .\label{}
\end{align}
The integral exists for $k>-1$ and may be evaluated to give the relation
\begin{equation}
\int_{0}^{\infty}\sum_{n=0}^{\infty} \frac{\operatorname{H}_{2n+1}^2(\rho)}{(2n+1)!} \frac{u^{2n+1}}{2^{2n+1}} \exp(-\rho^2) \rho^{2k-1} {\text{d}}\rho = \frac{1}{4} \frac{(k-1)!}{\sqrt{1-u^2}} \Big[\Big(\frac{1+u}{1-u}\Big)^k - \Big(\frac{1-u}{1+u}\Big)^k\Big] .\label{eq:I1}
\end{equation}
For $k=1$ the right hand side of the above expression may be evaluated explicitly and comparison of the $u^{2n+1}$ terms leads to 
\begin{equation}
\int_{0}^{\infty} \frac{\operatorname{H}_{2n+1}^2(\rho)}{(2n+1)!} \exp(-\rho^2) \rho {\text{d}}\rho = 2 \Big(\frac{(2n+1)!}{n!}\Big)^2 , \label{}
\end{equation} 
in agreement with (\ref{eq:M3}). 

We now return to the study of the recursion relations governing the matrix elements of the general $|nl\rangle$ states of the 3-d oscillator. The relation given in (\ref{eq:P0}) in general relates three different matrix elements. However, for the oscillator potential, for the special values $j=0,-1,1,-2l$ and $2l+2$ relations between just two matrix elements arise. As noted earlier in the general discussion following (\ref{eq:N12}) $j=1$ gives the virial theorem and the case $j=0$ yields (\ref{eq:M1}). Additional relations to note are
\begin{alignat}{2} 
l\ge 1  , j=-2l  &\Longrightarrow & 8l\epsilon \left\langle nl\middle|\rho^{-2l-1}\middle|nl\right\rangle &= (4l-2) \left\langle nl\middle|\rho^{1-2l}\middle|nl\right\rangle + C_{nl}^2 (2l+1)^2 \label{eq:M0}\\
l\ge 1 ,  j=-1 &\Longrightarrow & \left\langle nl\middle|\frac{1}{\rho^2}\middle|nl\right\rangle &= \frac{(2l-1)(2l+3)}{4\epsilon} \left\langle nl\middle|\frac{1}{\rho^4}\middle|nl\right\rangle ,  \label{}
\end{alignat}

For $j>0$ the relation given in (\ref{eq:P0}) leads to
\begin{equation}
\left\langle \rho^{j+1} \right\rangle = \frac{2j}{j+1} \epsilon \left\langle\rho^{j-1} \right\rangle - \frac{(j-1)}{4(j+1)} (2l+j) (2l+2-j)\left\langle\rho^{j-3}\right\rangle , j=1,2,\dots . \label{eq:M4}
\end{equation}
The recursion relation given in (\ref{eq:M4}) breaks up into two different relations one of which connects the matrix elements of even powers of $\rho$ and another which connects the matrix elements of odd powers of $\rho$.

Using $j=2k+1$ a recursion relation for the expectation values of even powers of $\rho$ may be found which can be expressed as a recursion relation for the expectation values of powers of $v$ in the form
\begin{equation}
\left\langle v^{k+1} \right\rangle = \Big(\frac{2k+1}{2k+2}\Big) \epsilon \left\langle v^k \right\rangle - \frac{k}{16(k+1)} (2l+1+2k) (2l+1-2k) \left\langle v^{k-1} \right\rangle\ . \label{eq:M5}
\end{equation}
The first few iterations yield
\begin{align}
\left\langle v \right\rangle &=  \frac{1}{2} \epsilon \quad   {\hbox{where}}\quad  \epsilon \equiv \epsilon_{nl} = \frac{1}{2} (4n + l +3) \label{eq:M6}\\
\left\langle v^2 \right\rangle &=  \frac{3}{4} \epsilon \left\langle v \right\rangle - \frac{1}{32} (2l + 3) (2l - 1) = \frac{3}{8} \epsilon^2 - \frac{1}{32} (2l+3)(2l-1) \label{eq:M7}\\
\left\langle v^3 \right\rangle &=  \frac{5}{6} \epsilon \left\langle v^2 \right\rangle - \frac{1}{24} (2l +5 ) (2l - 3) \left\langle v \right\rangle = \frac{5}{16} \epsilon^3 - \frac{\epsilon}{64} \big[12l(l+1) - 25\big] . \label{eq:M8}
\end{align}
The recursive scheme given in (\ref{eq:M5}) enables the evaluation $\left\langle v^j \right\rangle$ for any positive value of $j$ in terms of $\epsilon$ and $l$. 

For $l=0$ starting from $\left\langle\rho\right\rangle$ given in (\ref{eq:M2}) and (\ref{eq:M3}) the expectation values of all odd powers of $\rho$ may be found using (\ref{eq:M4}) recursively. For $l\ne 0$, (\ref{eq:M0}) can be used as the starting point of a recursive scheme and  $C_{nl}^2$ and $\left\langle\rho^{1-2l}\right\rangle$ may be used to find the expectation values of higher odd powers of $\rho$.

It follows from (\ref{eq:M6}) that for the oscillator $\left\langle T\right\rangle=\left\langle v\right\rangle =\epsilon/2 $ which is a well known result. This result taken together with (\ref{eq:g6}) shows that $\left\langle T^2\right\rangle = \left\langle v^2\right\rangle$ . Furthermore (\ref{eq:g7}) and (\ref{eq:M6})--(\ref{eq:M8}) may be used to establish that $\left\langle T^3\right\rangle =\left\langle v^3\right\rangle$. Similarly (\ref{eq:M5}) and (\ref{eq:g8}) may be used to show that
\begin{align}
\left\langle v^4\right\rangle &= \frac{7}{8} \epsilon \left\langle v^3\right\rangle - \frac{3}{64} (2l+7)(2l-5) \left\langle v^2\right\rangle \\
&= \frac{35}{128} \epsilon^4 - \frac{5\epsilon^2}{256} \big(12 l(l+1) - 49\big) + \frac{3}{2048}(2l+7)(2l-5)(2l+3)(2l-1) \\
\left\langle T^4\right\rangle - \left\langle v^4\right\rangle &= \epsilon^4 - 4\epsilon^3 \left\langle v\right\rangle +6\epsilon^2 \left\langle v^2\right\rangle - 4\epsilon\left\langle v^3\right\rangle + 8\epsilon\left\langle v\right\rangle - 8\left\langle v^2\right\rangle + \frac{3}{4} - l(l+1) = 0 . \label{}
\end{align}
It would seem to be the case that $\left\langle T^n\right\rangle=\left\langle v^n\right\rangle$ for all positive integer values of $n$ which is not so well known but not an altogether surprising result since the Hamiltonian is a symmetric function of the position and momentum variables.

\subsection{Linear potential}
For the linear potential $v=\rho/2$ the relation given in (\ref{eq:P0}) leads to
\begin{equation}
1 =   |\Psi_{nl}(0)|^2 \delta_{0,l} + 2 \left\langle nl\middle| \frac{l(l+1)}{\rho^3} \middle| nl \right\rangle \label{eq:L1}
\end{equation}
For $l=0$ the radial Sch\"{o}rdinger equation 
\begin{equation}
\Big(-\frac{\text{d}^2}{\text{d}\rho^2} + \rho \Big) R_{n0} = 2\epsilon  R_{n0}, \qquad Z\equiv \rho-2\epsilon  \Longrightarrow \frac{\text{d}^2R_{n0}}{\text{d} Z^2} = Z R_{n0} \label{eq:L2}
\end{equation}
may be identified as the equation satisfied by Airy functions. The energy eigenvalues can be identified as $ E_n= -Z_n/2$ and  the  radial eigenfunctions $R_{n0}$ are proportional to the Airy functions $\text{Ai}(\rho+Z_n)$ where $Z_n$ are the zeros of Airy functions. It may be shown using (\ref{eq:L2}) and the boundary condition on $R_{n0}$ as $\rho\rightarrow\infty$  that
\begin{align}
\big(R^{\prime}(Z)\big)^2 &= \int 2 R^\prime R^{\prime\prime} {\text{d}}Z = 2\int ZR R^\prime = ZR^2 - \int R^2 {\text{d}}Z \notag\\
\int_0^\infty  \text{Ai}^2(\rho+Z_n) {\text{d}}\rho &= \int_{Z_n}^\infty \text{Ai}^2(Z) {\text{d}}Z = \big(\text{Ai}^\prime(Z_n)\big)^2  ,\label{}
\end{align}
where a prime denotes a derivative with respect to $Z$. The normalised radial eigenfunctions are given by
\begin{equation}
 R_{n0} = \frac{\text{Ai}(\rho+Z_n)}{\text{Ai}^\prime(Z_n)} , \label{eq:L3}
\end{equation}
which shows that the normalised spherically symmetric radial functions have unit slope at $\rho=0$ which is in agreement with the $l=0$ limit of (\ref{eq:L1}). 

Even though the eigenfunctions for $l\ge 1$ have to be found by numerical integration it is clear from (\ref{eq:L1}) that the expectation value of the centrifugal force must always have the value $1/2$. Using (\ref{eq:P0}) it may be seen that the special values of $j$ for which the relation involves just two matrix elements are $j=-1/2,0,1,2l$ and $2l+2$. As noted earlier in eq.~(\ref{eq:N12}) $j=1$ gives rise to the virial theorem. Additional relations to note are
\begin{align}
 l\ge 1  ,  j=0  \Longrightarrow \left\langle\frac{l(l+1)}{\rho^3}\right\rangle &= \frac{1}{2}\\
 l\ge 1  ,  j= -\frac{1}{2} \Longrightarrow  \left\langle\rho^{-7/2}\right\rangle &= \frac{32}{3} \frac{\epsilon}{(4l-1)(4l+5)} \left\langle\rho^{-3/2}\right\rangle .  \label{}
\end{align}

For the linear potential the matrix element relation for all $l$ given in (\ref{eq:P0}) for  positive $j$ becomes
\begin{equation}
\left\langle \rho^j\right\rangle = \frac{4j}{2j+1} \epsilon \left\langle\rho^{j-1}\right\rangle - \frac{j-1}{2(2j+1)} (2l+j) (2l+2-j) \left\langle\rho^{j-3}\right\rangle .\label{eq:L4}
\end{equation}
The choice $j=1$ leads to the virial theorem
\begin{equation}
\left\langle v\right\rangle = \left\langle \frac{\rho}{2}\right\rangle = \frac{2\epsilon}{3} = 2\left\langle T\right\rangle .\label{}
\end{equation}
For $l=0$ all the expectation values of positive powers of $v$ may be found in closed form using (\ref{eq:L4}) in the form:
\begin{equation}
\left\langle n0|v^j|n0\right\rangle = \frac{2j}{2j+1} \epsilon \left\langle n0|v^{j-1}|n0\right\rangle + \frac{j(j-1)(j-2)}{16(2j+1)} \left\langle n0|v^{j-3}|n0\right\rangle ,\label{}
\end{equation}
and the first few iterations yield
\begin{align}
\left\langle n0|v^2|n0\right\rangle &= \left\langle n0|\frac{\rho^2}{4}|n0\right\rangle = \frac{4}{5} \frac{2}{3} \epsilon^2 \\
\left\langle n0|v^3|n0\right\rangle &= \left\langle n0|\frac{\rho^3}{8}|n0\right\rangle = \frac{6}{7} \frac{4}{5} \frac{2}{3} \epsilon^3 + \frac{3}{56} \\
\left\langle n0|v^4|n0\right\rangle &= \left\langle n0|\frac{\rho^4}{16}|n0\right\rangle = \frac{8}{9} \frac{6}{7} \frac{4}{5} \frac{2}{3} \epsilon^4 + \frac{1}{9} \Big(1 + \frac{3}{7}\Big) \epsilon .\label{}
\end{align}

For $l=1$ knowledge of $\epsilon_{n,1}$ and $\left\langle n1|\rho^{-1}|n1\right\rangle$ enables the determination of $\left\langle\rho^j\right\rangle, j>0$, through repeated application of (\ref{eq:L4}). Similar statements apply for $l>1$.

\subsection{Coulomb potential}
For the attractive Coulomb potential $v=-\rho^{-1}$ the relation given in (\ref{eq:P0}) leads to
\begin{equation}
\left\langle nl| \rho^{-2} |nl\right\rangle =   \frac{1}{2} C_{n0}^2 \delta_{0,l} +  l(l+1) \left\langle nl| \rho^{-3} | nl \right\rangle  .\label{eq:C1}
\end{equation}
This relationship is easy to verify for the state with $n=l+1$ for which the eigenfunction is nodeless:
\begin{align}
R_{l+1,l} &= C_{l+1,l} \rho^{l+1} \exp \Big(-\frac{\rho}{l+1}\Big) \\
\frac{\left\langle l+1,l| \rho^{-2} |l+1,l\right\rangle}{\left\langle l+1,l| \rho^{-3} |l+1,l\right\rangle} &= \frac{(2l)!}{(2l-1)!} \frac{l+1}{2} = l(l+1),   l\ne 0 \\
\left\langle 1,0| \rho^{-2} |1,0\right\rangle &= 2  ,  C_{1,0}^2 = \Big( \int_0^\infty \rho^2 \exp (-2\rho) {\text{d}}\rho\Big)^{-1} = 4 ,\label{}
\end{align}
in agreement with (\ref{eq:C1}). For the spherically symmetric states with arbitrary $n$ it can be verified using the Coulomb wavefunctions (Landau and Lifshitz 119-122) that
\begin{equation}
\left\langle n0| \rho^{-2} |n0\right\rangle = \frac{2}{n^3} = \frac{C_{n0}^2}{2} . \label{}
\end{equation}

Using the relation given in (\ref{eq:P0}) it may be seen that the special values of $j$ for which the relation involves just two matrix elements are $j=0,1/2,1,2l$ and $2l+2$. As noted earlier in (\ref{eq:N12}) the case $j=1$ yields the virial theorem. Additional relations to note are  
\begin{align}
l\ge 1  ,  j=0  \Longrightarrow \left\langle\frac{l(l+1)}{\rho^3}\right\rangle &= \left\langle\frac{1}{\rho^2}\right\rangle\\
l\ge 1  ,  j=\frac{1}{2}  \Longrightarrow \left\langle\rho^{-\frac{5}{2}}\right\rangle &= -\frac{32\epsilon}{(4l+1)(4l+3)} \left\langle\rho^{-\frac{1}{2}}\right\rangle . \label{}
\end{align}

For positive definite values of $j$ (\ref{eq:P0}) leads to the relation
\begin{equation}
2j \epsilon \left\langle\rho^{j-1}\right\rangle + (2j-1) \left\langle\rho^{j-2}\right\rangle - \frac{(j-1)}{4} (2l+j) (2l+2-j) \left\langle\rho^{j-3}\right\rangle  = 0 , \label{}
\end{equation}
which is Kramer's relation. Various Coulomb matrix elements can be deduced as functions of $E$ and $l$. $j=1$ is the virial theorem giving $\left\langle v\right\rangle$ in terms of $\epsilon$, $j=2$ gives an expression for $\left\langle \rho\right\rangle$, $j=3$ gives $\left\langle \rho^2\right\rangle$ { etc.} Thus $\left\langle \rho^j \right\rangle $ for all positive values of  $j$ can be found as functions of $E$ without the need to calculate even a single matrix element.

\section{Generalization of virial theorems in Classical Mechanics}

The Hamiltonian in Classical Mechanics $H$ and a generating function $G$ given by
\begin{equation}
H = \sum_j \frac{p_j^2}{2M} + V(r_j) ,  G = \sum_j p_j r_j ,\label{eq:V1}
\end{equation}
may be used to establish the time evolution of $G$ using Poisson brackets in the form
\begin{equation}
\frac{{\text{d}}G}{{\text{d}}t} = -\{H,G\} = -\sum_j\Big( \frac{\partial H}{\partial r_j} \frac{\partial G}{\partial p_j} - \frac{\partial H}{\partial p_j} \frac{\partial G}{\partial r_j}\Big) = \sum_j\Big(\frac{p_j}{M} p_j - \frac{\partial V}{\partial j} r_j\Big) = 2T - {\bf r}.{\bf \nabla}V  .\label{}
\end{equation}
Time averaging over a period of a periodic orbit gives the standard virial relation
\begin{equation}
2\left\langle T\right\rangle - \left\langle {\bf r}.{\bf \nabla}V \right\rangle = 0 ,\label{}
\end{equation}
where the angular brackets now represent time averages. The corresponding equation for the time evolution of the function $GF(r)$ may be found in the form
\begin{align}
\frac{\partial}{\partial t} (GF) &= -\{H,GF\} = -\{H,G\} F - G\{H,F\} \notag\\
&=  \big(2T - {\bf r}.{\bf \nabla}V\big) F + \frac{G}{M} \Big(\sum_jp_j r_j\Big) \Big(\frac{1}{r}\frac{\partial F}{\partial r}\Big) =  \big(2T - {\bf r}.{\bf \nabla}V\big) F + \frac{G^2}{M}  \frac{1}{r}\frac{\partial F}{\partial r} .\label{eq:V2}
\end{align}
Time averaging over a period of a periodic orbit then gives the generalized relation
\begin{equation}
\left\langle\big(2T - {\bf r}.{\bf \nabla}V\big) F \right\rangle + \frac{1}{M} \left\langle G^2  \frac{1}{r}\frac{\partial F}{\partial r}\right\rangle = 0 . \label{}
\end {equation}
In $3$ dimensions $G^2$ may be related to the square of the angular momentum using
\begin{equation}
G^2 = \big(\sum_j p_j r_j\big)^2 = ({\bf r.p})^2 =  {\bf r.r} {\bf p.p} - ({\bf r }\times{\bf p}).({\bf r}\times{\bf p}) = r^2 p^2 - l^2 .\label{}
\end{equation}
For a spherically symmetric potential in $3$ dimensions we can thus establish a generalization of the virial theorem in the form
\begin{equation}
\left\langle 2T\Big(F + r \frac{\partial F}{\partial r}\Big)\right\rangle - \left\langle F r \frac{\partial V}{\partial r} + \frac{l^2}{Mr^2} r \frac{\partial F}{\partial r}\right\rangle = 0 .\label{}
\end{equation}
This expression acquires a simple form when expressed in terms of $f\equiv r F$:
\begin{equation}
\left\langle 2T \frac{\partial f}{\partial r}\right\rangle = \left\langle \frac{\partial V}{\partial r}\right\rangle + \frac{l^2}{M} \left\langle \frac{1}{r} \frac{\partial}{\partial r} \frac{f}{r}\right\rangle . \label{}
\end{equation}
The standard virial theorem corresponds to the choice $f=r$. For a general function $f$ the generalization of the virial theorem may also be given in the form
\begin{equation}
 \left\langle \frac{1}{f} \frac{\partial}{\partial r} (f^2 T_r) \right\rangle = 0   {\hbox{where}}  T_r = E - V - \frac{l^2}{2Mr^2} ,\label{eq:V6}
\end{equation}
is the radial kinetic energy. The classical time average may be compared to the expectation values in Quantum Mechanics with the replacement of $l^2$ by the eigenvalue $l(l+1)$. The quantity $Q$ considered in the Quantum case is related to the radial kinetic energy by $Q = -2T_r$. These identifications enable the comparison of the classical result in (\ref{eq:V6}) with the quantum result given in (\ref{eq:N7}) which may be given in the form
\begin{equation}
\left\langle \frac{1}{f} \frac{\partial}{\partial \rho} (f^2 Q) \right\rangle = \frac{1}{2} \left\langle\frac{\partial^3f}{\partial \rho^3}\right\rangle , \label{eq:V5}
\end{equation}
for any $f(\rho)$ which for small $\rho$ tends to $\rho^j, j>0$. It is evident that there is agreement between the two results for the choices $j=0,1,2$ but there is an additional term in the quantum case for other choices of $f$ for which the third derivative of $f$ is non-vanishing. The discrepancy between the two results may be traced to the non-vanishing commutation relation between $p$ and $r$ which implies that in general ${\bf p.r \ne r.p}$ and consequently $GF \ne FG$. These distinctions must be taken into account in carrying over the analysis based on the generating function $G$ in Classical Mechanics to Quantum Mechanics.

\subsection{Quantum mechanics}

 The Hamiltonian $H$ in Quantum Mechanics is obtained from (\ref{eq:V1}) by the replacement $p\rightarrow -{\text{i}}\hbar\nabla$ and $G$ is replaced by a hermitian operator constructed using a symmetric average of ${\bf r.p}$ and ${\bf p.r}$. The Poisson brackets become commutators and $[A,BC] = [A,B]C + B[A,C], [AB,C] = A[B,C]  + [A,C]B$. If $\phi$ is a non-degenerate eigenstate of $H$ then 
\begin{equation}
\left\langle \phi \middle| [H,A]\middle| \phi\right\rangle = 0  ,\label{eq:V7}
\end{equation}
for any 'reasonable' operator $A$. These relations may be used to show that
\begin{align}
G_q &= \frac{{\bf r.p + p.r}}{2} \notag \\
[H,G_q] &= \Big[\Big(\frac{p^2}{2M} + V\Big), \frac{{\bf r.p+p.r}}{2} \Big] = -{\text{i}}\hbar \Big(\frac{p^2}{2M} - {\bf r.\nabla}V \Big) \label{eq:V8} \\
\left\langle\phi | [H,G_q] |\phi\right\rangle &= 0    \Longrightarrow   \left\langle \phi | (2T - {\bf r.\nabla}V ) | \phi\right\rangle  = 0 ,\label{eq:V9}
\end{align}
which is the virial theorem in Quantum Mechanics. Similarly for any function $F(r)$ the commutator with $F$ may be evaluated to give
\begin{equation}
[H,F] = \frac{1}{2M} [p^2,F] = -\frac{{\text{i}}\hbar}{2M} \big( {\bf p.\nabla}F + {\bf \nabla}F.{\bf p}\big) = -\frac{\hbar^2}{2M}\nabla^2F - \frac{{\text{i}}\hbar}{m}{\bf \nabla}F.{\bf p}  .\label{eq:V10}
\end{equation}
The results given in (\ref{eq:V8}) and (\ref{eq:V10}) may be used to establish that
\begin{align}
\big[H, (F G_q + G_q F)] &= \Big([H,F] G_q + G_q [H,F] + F [H,G_q] + [H,G_q] F \Big) \notag\\
\frac{iM}{\hbar}\big[H, (F G_q + G_q F)] &= \big({\bf p.\nabla}F G_q + G_q {\bf p.\nabla}F\big) + \big({\bf \nabla}F.{\bf p} G_q + G_q {\bf \nabla}F.{\bf p} \big) \notag\\
 &+ \big(p^2 F + F p^2\big) - M \big({\bf r.\nabla}V F + F {\bf r.\nabla}V\big) . \label{eq:V11}
\end{align}
To make further progress we use the following relations valid in $3$ dimensions
\begin{align}
{\bf r.p} &= -{\text{i}}\hbar r \frac{\partial}{\partial r},  {\bf p.r} = -{\text{i}}\hbar \Big(r \frac{\partial}{\partial r} + 3\Big)  ,  {\bf r.p - p.r}  = 3 {\text{i}}\hbar \label{eq:V12}\\
\frac{\partial^2}{\partial r^2} &= \nabla^2 - \frac{2}{r} \frac{\partial}{\partial r} - \frac{l(l+1)}{r^2} ,  T_r = T - \frac{l(l+1)}{2Mr^2}  ,  T = E - V = -\frac{\hbar^2}{2M}\nabla^2 . \label{}
\end{align}
Using (\ref{eq:V7}), (\ref{eq:V11}) and (\ref{eq:V12}) and the Schr\"{o}dinger equation satisfied by $\phi$ it can be established after some algebra that
\begin{align}
\left\langle\phi\middle| \Big[\big(2T - {\bf r.\nabla}V \big) F + 2{\bf r.\nabla}F \Big(T - \frac{l(l+1)}{2Mr^2} \Big)\Big] \middle|\phi\right\rangle &= -\left\langle\phi\middle|  \frac{\hbar^2}{4M} \frac{\partial^3}{\partial r^3} (rF) \middle|\phi\right\rangle \\
f\equiv r F \Longrightarrow \left\langle\phi \middle| \Big(\frac{1}{f} \frac{\partial}{\partial r}(f^2T_r) \Big) \middle| \phi\right\rangle &= -\left\langle \phi\middle| \frac{\hbar^2}{4M} \frac{\partial^3f}{\partial r^3}\middle| \phi\right\rangle  ,\label{eq:V13}
\end{align}
reproducing the result given in (\ref{eq:V5}). The term on the right hand side of (\ref{eq:V13}) is non-vanishing except when $f=r^0$ or $r$ or $r^2$. Comparison of (\ref{eq:V6}) and (\ref{eq:V13}) shows that the presence of the term on the right hand side of (\ref{eq:V13}) is attributable to non-classical effects.

\section{Generalized virial theorems in $\protect\boldsymbol{N}$ dimensions}

The results presented in this work can be extended to arbitrary number of dimensions. The $N$ dimensional Laplacian is an operator function of a radial coordinate $r$ and $(N-1)$ angular coordinates $\theta_j, j=1,2,\dots ,N-1$ and can be given in the form
\begin{align}
\nabla_N^2 &= \frac{\partial^2}{\partial \rho^2} + \frac{N-1}{\rho} \frac{\partial}{\partial\rho} + \frac{K_{N-1}}{\rho^2} \\
&= \rho^{\frac{1-N}{2}} \frac{\partial^2}{\partial \rho^2} \rho^{(N-1)/2} + \frac{1}{\rho^2} \Big( K_{N-1} - \frac{(N-1)(N-3)}{4}\Big) , \label{}
\end{align}
where $K_{N-1}$ is the generalization of the angular momentum operator $L^2$ in $3$ dimensions to $N$ dimensions. (See, for example,  Gallup 1959 and the references cited there and, Hochstadt 1987). For potentials which are spherically symmetric in a $N$ dimensions the $N$ dimensional Schr\"{o}dinger equation for the wavefunction $\Psi$ can be converted to a radial equation for a suitably defined radial wavefunction $R$ as follows:
\begin{align}
\Psi(\rho,\theta_1,\dots,\theta_{N-1}) &= \psi(\rho) \Omega(\theta_1,\dots\theta_{N-1})\\
K_{N-1} \Omega &= -l_1 (l_1 + N-2) \Omega  ,  N>1\\
R \equiv \rho^{\frac{N-1}{2}} \psi &\Longrightarrow  \frac{\partial^2 R}{\partial \rho^2} = Q_N  R\\
K \equiv l_1 + \frac{N-1}{2} &\Longrightarrow Q_N = \frac{2M}{\hbar^2} \big( v - \epsilon\big) + \frac{K(K-1)}{\rho^2}  ,\label{}
\end{align}
where $\Omega$ are generalized spherical harmonics in $N$ dimensions. The expectation values may as before be defined in terms of a weighted integral using a probability distribution defined by $P=R^2$ where $P$ is a solution of
 \begin{equation}
 {\dddot P} - 4Q_N{\dot P} -  2{\dot Q_N} P  = 0  .\label{}
\end{equation} 
The generalized virial theorem in $N$ dimensions may be given in the form
\begin{align}
\left\langle\frac{1}{f} \frac{\partial(f^2 Q_N)}{\partial\rho}\right\rangle &= \frac{1}{2} \left\langle\frac{\partial^3 f}{\partial\rho^3}\right\rangle + \Delta_N \\
2\Delta_N &= \big( f{\ddot P} - {\dot f} {\dot P} + {\ddot f} P - 4Q_N f P \big)|_{\rho=0} \\
&= B C \Big(2(2l_1 + N - 2)^2-(2N-3)(N-3)\Big) \delta_{q,-2\lambda-N+3} \\
q &= \frac{\partial \ln f}{\partial \ln \rho} |_{\rho=0}   ,  C = \frac{P}{\rho^{2l_1 +N-1}} |_{\rho=0}  ,  B = \frac{f}{\rho^q} |_{\rho=0}  .\label{}
\end{align}
Another equivalent version of the generalized virial theorem is
\begin{align}
q_0 \equiv & -(2\lambda + N - 3) \\
q> q_0 \Longrightarrow &\left\langle\frac{1}{f} \frac{\partial(f^2 Q_N)}{\partial\rho}\right\rangle = \frac{1}{2} \left\langle\frac{\partial^3 f}{\partial\rho^3}\right\rangle \\
q=q_0 \Longrightarrow & \left\langle\frac{1}{f} \frac{\partial}{\partial\rho}(f^2 Q_N)\right\rangle - \frac{1}{2} \left\langle\frac{\partial^3 f}{\partial\rho^3}\right\rangle = B  C \Big(2 (q_0-1)^2-(2N-3)(N-3)\Big) . \label{Eq:G4}
\end{align}
Correspondence with the results for $N=1,2,3$ can be made with the identification that $l_1=0$ if $N=1$, $l_1=l$ if $N=2 ,3$ and
\begin{align}
Q_1 &= \frac{2M}{\hbar^2} \big(v - \epsilon\big) \\
Q_2 &= \frac{2M}{\hbar^2} \big(v - \epsilon\big)  + \frac{(4l^2 - 1)}{4 \rho^2}\\
Q_3 &= \frac{2M}{\hbar^2} \big(v - \epsilon\big)  + \frac{l(l+1)}{\rho^2} . \label{Eq:G5}
\end{align}

The generalization to $N$ dimensions can also be carried out in Classical Mechanics and the result may be given in the form
\begin{equation}
\left\langle\frac{1}{f} \frac{\partial}{\partial\rho}(f^2 Q_{\text{c}})\right\rangle = 0  ,  {\hbox{where}}  Q_{\text{c}} = (v - \epsilon) +\frac{1}{2M r^2} \sum_{i=1}^N \sum_{j>i}^N ( x_ip_j - x_j p_i)^2  ,\label{}
\end{equation}
where the angular brackets now refer to time averages for periodic orbits. It is evident that the classical and quantum results differ in the following three aspects:

${\bullet}$ The value for the square of the generalized angular momentum in $N$ dimensions in Classical Mechanics must be replaced by the eigenvalues of the generalized angular momentum operator in Quantum Mechanics.

${\bullet}$ The classical result does not contain any analogue of the expectation value of the third derivative of $f$.

${\bullet}$ The classical result does not have the analogue of $\Delta_N$ which gives rise to an additional term when $f$ has specific forms as $\rho \rightarrow 0$. This additional term has its origins in the boundary terms arising from the integration by parts in the evaluation of the quantum mechanical expectation values and has no parallels in Classical Mechanics.

All these three differences may ultimately be traced to the existence of non-vanishing commutation relations between coordinates and momenta in Quantum Mechanics.

\section{ APPENDIX: Angular momentum in $N$ dimensions}

In N dimensions the Cartesian coordinates $[x_j,j=1,2,\dots,N]$ may be related to $N$-dimensional spherical coordinates $[r,\theta_j, j=1,2,\dots,N-1]$ by the mapping
\begin{align}
x_1 &= r \cos\theta_1  ,  x_2 = r \sin\theta_1 \cos\theta_2  ,  x_3 = r\sin\theta_1 \sin\theta_2 \cos\theta_3 ,\dots \\
x_{N-1} &= r \Big(\prod_{j=1}^{N-2} {\sin\theta_j}\Big) \cos\theta_{N-1}  ,  x_{N} = r \Big(\prod_{j=1}^{N-2}{ \sin\theta_j}\Big) \sin\theta_{N-1}\\
0&\leq \theta_{N-1} \leq 2\pi  ,  0\leq \theta_{j} \leq \pi , j=1,2,\dots ,N-2  .  \label{}
\end{align}
The inverse transformation expressing the spherical coordinates in terms of the Cartesian coordinates may also be identified using the Pythagoras theorem and inverse tangent functions in a manner similar to that in $3$ dimensions:
\begin{equation}
r^2 = \sum_{j=1}^N x_j^2  , \theta_{N-1} = \tan^{-1}\Big(\frac{x_N}{x_{N-1}}\Big)  , \theta_{N-2} = \tan^{-1} \Big(\frac{\sqrt{x_N^2+x_{N-1}^2}}{x_{N-2}}\Big) ,\dots . \label{}
\end{equation}
The differential elements may be identified as
\begin{equation}
\text{d}r , r \text{d}\theta_1 , r \sin\theta_1 \text{d}\theta_2,\dots , r \Big(\prod_{j=1}^{N-2} \sin\theta_j\Big) \text{d}\theta_{N-1} . \label{}
\end{equation}

The kinetic energy in Classical Mechanics may be given in the form
\begin{align}
T &= \frac{1}{2} M \Big(\frac{\partial r}{\partial t}\Big)^2 + \frac{L^2}{2Mr^2} \\
L^2 &= \big(Mr^2\big)^2 \Big[\Big(\frac{\partial \theta_1}{\partial t}\Big)^2 + \sum_{j=2}^{N-1} \Big(\prod_{k=2}^j {\sin^2\theta_k}\Big) \Big(\frac{\partial\theta_j}{\partial t}\Big)^2\Big] . \label{}
\end{align}

In Quantum Mechanics coordinates and momenta become operators and the kinetic energy is represented in terms of the Laplacian operator. The Laplacian in $N$ dimensions may be given in the form
\begin{align}
\nabla_N^2 &= \frac{1}{r^{N-1}} \frac{\partial}{\partial r} r^{N-1} \frac{\partial}{\partial r} + \frac{1}{r^2}K_{N-1} \notag\\
K_{N-1} &= \frac{1}{\sin^{N-2}\theta_1} \frac{\partial}{\partial\theta_1} \sin^{N-2}\theta_1 \frac{\partial}{\partial\theta_1} + \frac{1}{\sin^2\theta_1}K_{N-2} \notag\\
K_{N-2} &= \frac{1}{\sin^{N-3}\theta_2} \frac{\partial}{\partial\theta_2} \sin^{N-3}\theta_2 \frac{\partial}{\partial\theta_2} + \frac{1}{\sin^2\theta_2}K_{N-3}  \notag\\
&\dots \notag\\
K_{2} &= \frac{1}{\sin\theta_{N-2}} \frac{\partial}{\partial\theta_{N-2}} \sin\theta_{N-2} \frac{\partial}{\partial\theta_{N-2}} + \frac{1}{\sin^2\theta_{N-2}}K_{1} \notag\\
K_1 &= \frac{\partial^2}{\partial\theta^2_{N-1}} . \label{}
\end{align}

It may be shown that periodic solutions for the angular part arise when the eigenvalues of the operators $K_j$ are of the form $-l(l+j-1), l=0,1,2..$. For spherically symmetrical potentials in $N$-dimensions the eigenfunctions of the Schr\"{o}dinger equation may be identified by one radial quantum number $n$ and $N-1$ quantum numbers $l_j, j=1,2,..,N-1$, associated with the angular eigenfunctions $\Omega(\theta_1,\theta_2,..,\theta_{N-1})$:
\begin{align}
\Psi(r,\theta_1,\dots ,\theta_{N-1}) &= \psi_n(r) \Omega_{l_1\dots ,l_{N-1}}(\theta_1,\theta_2,\dots ,\theta_{N-1}) \notag\\
K_{N-1}\Omega &= -l_1(l_1+N-2) \Omega \notag\\
K_{N-2}\Omega &= -l_2(l_2+N-3) \Omega\notag\\
&\dots \notag\\
K_2\Omega &= -l_{N-2}(l_{N-2}+1) \Omega \notag\\
K_1\Omega &=  -l_{N-1}^2 \Omega  .\label{}
\end{align}

We now show that the angular eigenfunctions may be identified as associated Gegenbauer polynomials. The angular functions are solutions to differential equations of the form
\begin{equation}
\Big(\frac{1}{\sin^{j}\theta} \frac{\partial}{\partial\theta} \sin^{j}\theta \frac{\partial}{\partial\theta} - \frac{1}{\sin^2\theta} l_{j-1}(l_{j-1}+j-1)\Big) F_{l_j} =  -l_j(l_j+j) F_{l_j} , \label{}
\end{equation}
which may be brought to an easy identifiable differential equation using the transformation $z\equiv\cos\theta$:
\begin{equation}
\Big(-(1-z^2) \frac{\partial^2}{\partial z^2} + (j+1) z \frac{\partial}{\partial z} + \frac{l_{j-1}(l_{j-1}+j-1)}{1-z^2}\Big)F_{l_j} =  l_j(l_j+j) F_{l_j} . \label{eq:G1}
\end{equation}
If $l_{j-1}=0$ and $l_j=$ an integer then the solutions to this equation may be identified as the Gegenbauer polynomilas $C_{l_j}^{(j/2)}$ which are also known as ultraspherical polynomials (Abramowitz and Stegun 1965). If $l_{j-1}=m$ where $m$ is a non-zero positive integer, $m\leq l_j$, then by analogy with the definition of the associated Legendre polynomials we can define the associated Gegenbauer polynomials and it may shown that the solutions to (\ref{eq:G1}) belong to the family
\begin{align}
F_{l,m}^{(j)}(z) \equiv (-1)^m (1-z^2)^{m/2} \frac{\partial^m}{\partial z^m} C_{l}^{(\frac{j}{2})}(z)& \\
\Big(-(1-z^2) \frac{\partial^2}{\partial z^2} + (j+1) z \frac{\partial}{\partial z} + \frac{m(m+j-1)}{1-z^2}\Big)F_{l,m}^{(j)} &= l(l+j) F_{l,m}^{(j)} . \label{}
\end{align}
The angular eigenfunctions in $N$ dimensional space may be given in the form
\begin{align}
\Omega_{l_1,..,l_{N-1}}(\theta_1,\theta_2,\dots ,\theta_{N-1}) &= \exp(\text{i}l_{N-1}\theta_{N-1}) \prod_{j=1}^{N-2} {F_{l_{N-1-j},l_{N-j}}^{(j)}(z_{N-1-j})} \\
z_j \equiv \cos\theta_j  &, l_{N-1} \leq l_{N-2} \leq l_{N-3} \dots  l_2 \leq  l_1  .\label{}
\end{align}

If we denote the eigenvalues of $K_{N-1}$ by $\lambda \equiv l_1(l_1+N-2)$ then the radial part of the $N$-dimensional Laplacian can be written in the form
\begin{equation}
\nabla_r^2 = \frac{\partial^2}{\partial r^2} + \frac{N-1}{r} \frac{\partial}{\partial r} + \frac{\lambda}{r^2} . \label{}
\end{equation}

The first few normalized angular functions in $4$ dimensions, for example, are given by
\begin{align}
\lambda = 0 \Longrightarrow   \Omega_{000} &= \frac{1}{\pi\sqrt{2}} \\
\lambda = 3 \Longrightarrow   \Omega_{100} &= \frac{\sqrt{2}}{\pi} \cos\theta_1 \\
\Omega_{110} &=  \frac{1}{\pi\sqrt{2}}  \sin\theta_1 \cos\theta_2 \\
\Omega_{11\pm 1} &= \frac{1}{2\pi} \sin\theta_1 \sin\theta_2 \exp(\pm i\theta_3) . \label{}
\end{align}

\section{References}

1. C.Quigg and J.L.Rosner 1979 {\it Physics Reports} {\bf 56} 168--235

2. A.Messiah 1966 {\it Quantum Mechanics Vol.}1 (North-Holland Publishing Company- Amsterdam : John Wiley \& sons inc.) 431

3. L.D.Landau and E.M.Lifshitz 1965 {\it Quantum Mechanics} (Pergamon press) 119--122

4. G.A.Gallup 1959 {\it Journal of Molecular Spectroscopy} {\bf 3} 673--682

5. H.Hochstadt 1987 {\it The functions of Mathematical Physics} (Dover Publications)

6. M.Abramowitz and I.Stegun 1965 {\it Handbook of Mathematical Functions} Dover: New York 781

\end{document}